%% file: icip_go_rev.tex
\documentclass[]{article}

\usepackage{spconf}
\usepackage{graphicx}
\usepackage{amssymb}
\usepackage{epstopdf}
\usepackage{subfigure}
\usepackage{amsthm}
\usepackage{amsmath}
\usepackage{color}
\usepackage{bm}
\usepackage{cite}
\usepackage{tikz}
\usepackage{hyperref}

\include{definitions}

\newtheorem{thm}{Theorem}

\newtheorem{lem}[thm]{Lemma}
\newtheorem{prop}[thm]{Proposition}
\newtheorem{defn}[thm]{Definition}

\usepackage{xcolor}
\usepackage{pgfplots} 
\pgfplotsset{compat=newest} 
\pgfplotsset{plot coordinates/math parser=false} 
\pgfplotsset{ 
  legend style =
  {font=\small\sffamily},
  label style = {font=\footnotesize\sffamily},
	tick label style = {font=\footnotesize}
}
\newlength\figureheight 
\newlength\figurewidth 
\title{Structured low-rank recovery of piecewise constant signals with performance guarantees}
\name{Greg Ongie$^{\star}$, Sampurna Biswas$^{\dagger}$, and Mathews Jacob$^{\dagger}$\thanks{This work is supported by grants NSF CCF-1116067,  ACS RSG-11-267-01-CCE, and ONR-N000141310202.}}%
\address{$^{\star}$ Department of Mathematics, University of Iowa, IA, USA. \\
		$^{\dagger}$ Department of Electrical and Computer Engineering, University of Iowa, IA, USA.}

\begin{document}
%
\maketitle
%

\begin{abstract}
We derive theoretical guarantees for the exact recovery of piecewise constant two-dimensional images from a minimal number of non-uniform Fourier samples using a convex matrix completion algorithm. We assume the discontinuities of the image are localized to the zero level-set of a bandlimited function, which induces certain linear dependencies in Fourier domain, such that a multifold Toeplitz matrix built from the Fourier data is known to be low-rank. The recovery algorithm arranges the known Fourier samples into the structured matrix then attempts recovery of the missing Fourier data by minimizing the nuclear norm subject to structure and data constraints. This work adapts results by Chen and Chi on the recovery of isolated Diracs via nuclear norm minimization of a similar multifold Hankel structure. We show that exact recovery is possible with high probability when the bandlimited function describing the edge set satisfies an incoherency property. Finally, we demonstrate the algorithm on the recovery of undersampled MRI data.
\end{abstract}
\begin{keywords}
Structured Low-Rank Matrix Completion, Annihilating Filter Method, Finite Rate of Innovation, Compressed Sensing, MRI.
\end{keywords}

\section{INTRODUCTION}
\label{sec:intro}
The recovery of a linear combination of exponentials from their few uniform samples is a classical problem in signal processing with extensive applications. Prony's method, or one of its robust variants, attempts to recover the signal by estimating an annihilating polynomial whose zeros correspond to the frequency of the exponentials. The finite rate of innovation (FRI) framework \cite{vetterli2002sampling} extended these methods to recover more general signals that reduce to a sparse linear combination of Dirac delta functions under an appropriate transformation (e.g. differential operators, convolution). Recently, several authors have further extended FRI methods to recover such signals from their non-uniform Fourier samples \cite{chen2014robust,ALOHAarxiv,ALOHAisbi,haldar2014low,sampta2015} by exploiting the low-rank structure of an enhanced matrix (e.g. Hankel matrix in 1-D). Performance guarantees do exist when the transform is an identity and when the Diracs are well-separated \cite{chen2014robust}. 

The above signal models have limited flexibility in exploiting the extensive additional structure present in many multidimensional imaging problems. Specifically, the edges in multidimensional images are connected and can be modeled as smooth curves or surfaces. We have recently introduced a novel framework to recover piecewise polynomial images, whose edges are localized to smooth curves, from their uniform \cite{isbi2015,siam} and non-uniform \cite{sampta2015} Fourier samples; this work generalizes a recent extension of the FRI framework to curves \cite{pan2013sampling}. We model the piecewise smooth signal as having partial derivatives that vanish outside the zero level-set of a bandlimited function. This relation translates to an annihilation condition involving the uniform Fourier samples of the partial derivatives, which can be compactly represented as the multiplication of a specific structured matrix with the Fourier coefficients of the bandlimited function. Our earlier work has shown that the structured matrix is low-rank, and we used this property to recover the signal from its non-uniform Fourier samples with good performance. Efficient algorithms that work on the original signal samples rather than the structured high-dimensional matrix also were introduced \cite{isbi2016}. We observe the signal models in \cite{chen2014robust,ALOHAarxiv,haldar2014low} do not include the class of signals considered in this work. 

The main focus of this work is to introduce theoretical guarantees on the recovery of piecewise constant signals, whose discontinuities are localized to zero level-sets of bandlimited functions, from non-uniform Fourier samples. Since such signals cannot be expressed as a finite linear combination of isolated Diracs, the recovery guarantees in \cite{chen2014robust} cannot be directly extended to our setting. Specifically, the theory in \cite{chen2014robust} relies heavily on a explicit factorization of the enhanced matrix (e.g Vandermonde factorization of a Hankel matrix in the 1-D case), which is only available 
when the number of discontinuities are finite and well separated. Instead, we give a new description of the row and column subspace of the structured matrix, which allow us to derive incoherence measures based solely on properties of the bandlimited function describing the edge set of the image.
\begin{figure}
\centering
\includegraphics[width=0.45\textwidth]{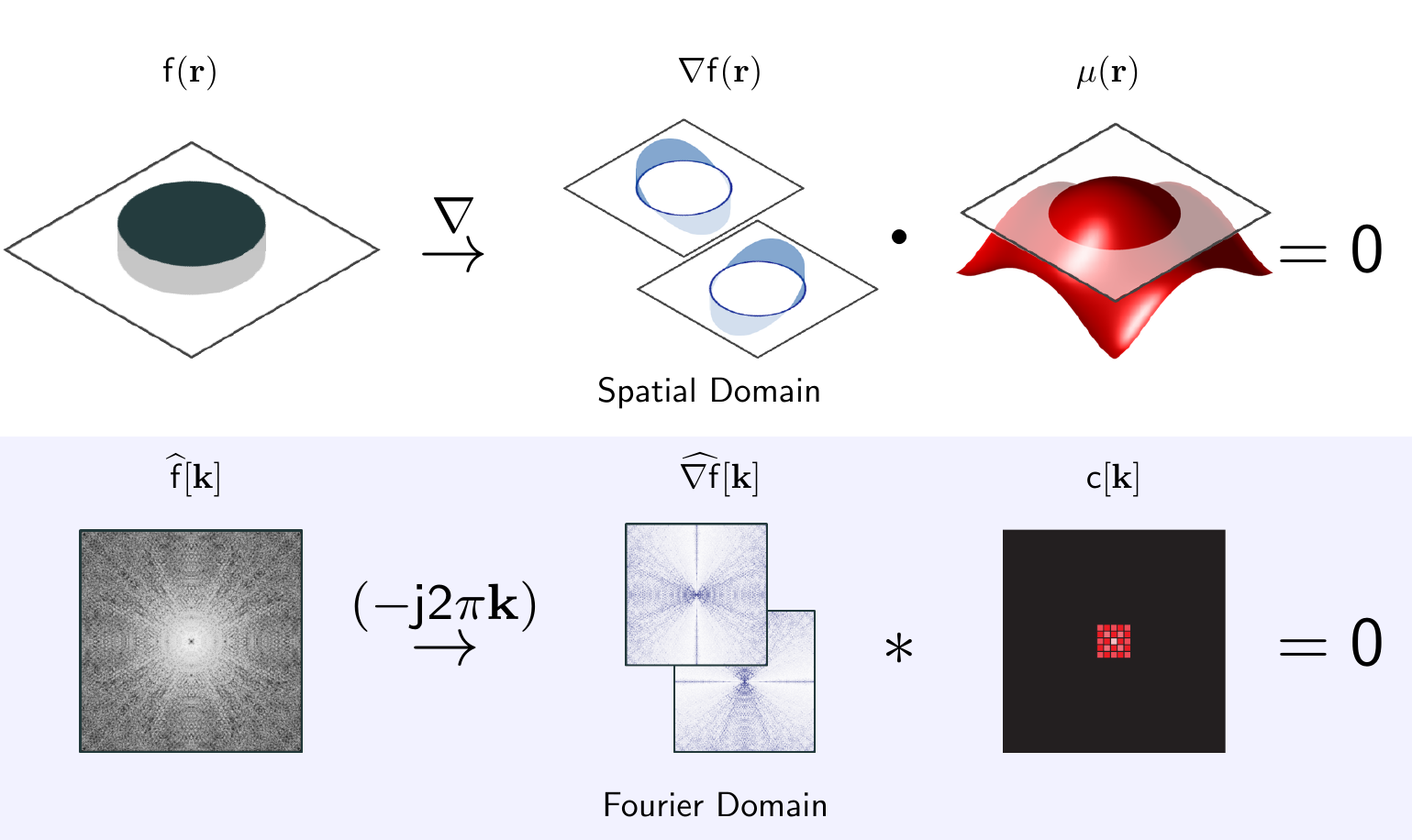}
\caption{\small Annihilation of a piecewise constant function in the spatial (top) and Fourier (bottom) domain.}
\label{illus}
\end{figure}
\section{Theory}
\subsection{Signal Model: 2-D Piecewise Constant Images}
We consider the recovery of a \emph{piecewise constant} function
\begin{equation}
\label{eq:pwc}
f(\mbf r) = \sum_{i=1}^N a_i ~\chi_{\Omega_i}(\mbf r),~~\text{ for all }~~\mbf r=(x,y) \in{[0,1]}^2,
\end{equation}
where $a_i \in \mathbb{C}$, and $\chi_{\Omega}$ denotes the characteristic function of the set $\Omega$. We assume the Fourier samples $\widehat{f}[\mbf k]$ specified by
\begin{equation}
\hat f[\mbf k] = \int_{[0,1]^2} f(\mbf r) e^{-j 2 \pi \mbf k\cdot \mbf r}; ~\mbf k \in \mathbb{Z}^2,
\end{equation}
are available at a subset of non-uniform locations $\mbf k \in \Theta$ belonging to a rectangular set of uniform sampling locations $\Gamma \subset \mathbb{Z}^2$ in Fourier domain. 

\begin{figure}[ht!]
\centering
\includegraphics[width=0.35\textwidth]{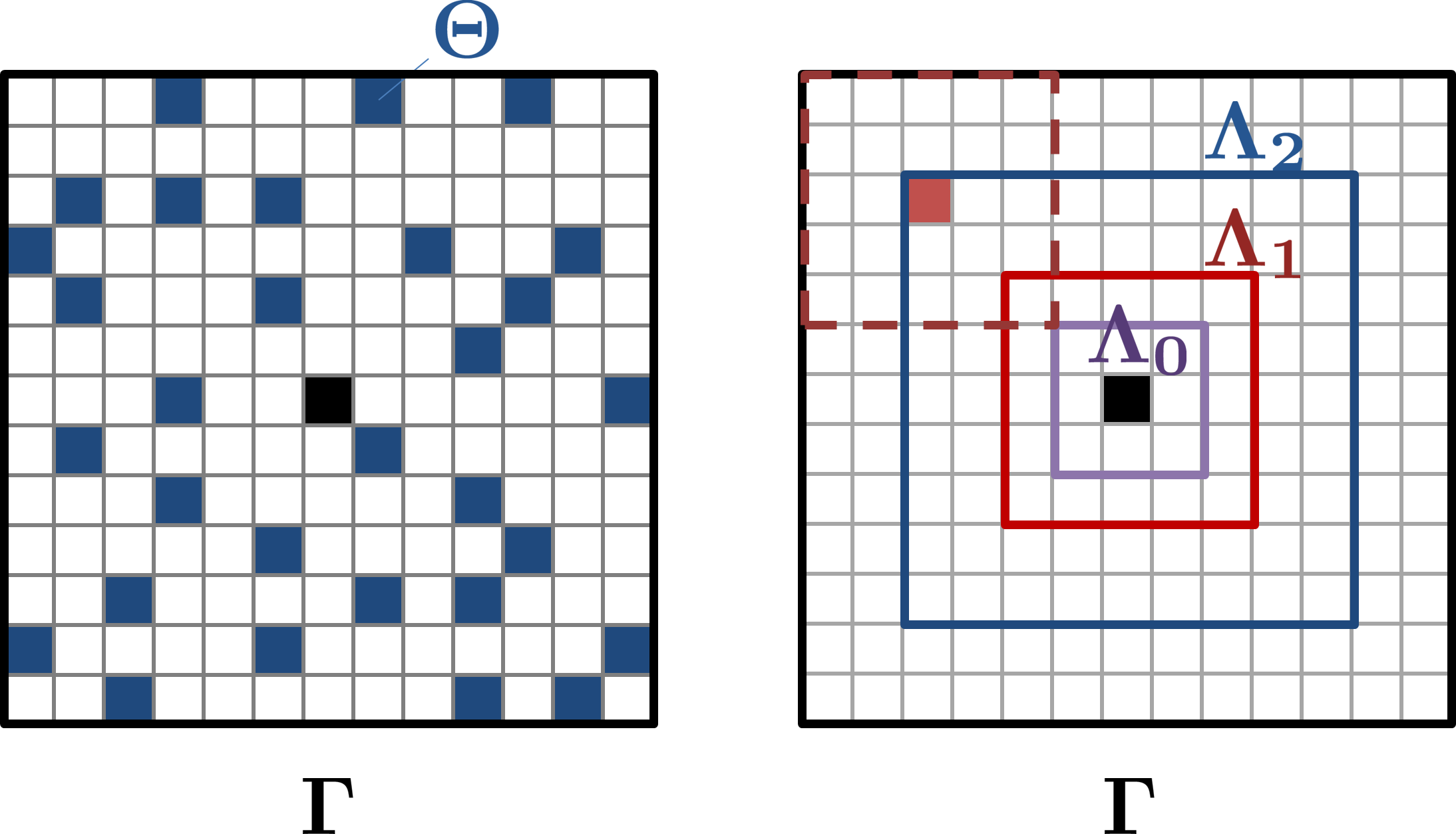}
\caption{\small Fourier domain support sets used in constructing the structured matrix $\mathcal{T}(\hat f)$. The grid $\Gamma$ represent a rectangular sampling window in $\mathbb{Z}^2$ (index $(0,0)$ marked in black). (left) $\Theta$ is a collection of uniform random sampling locations. (right) $\Lambda_0$ is the Fourier support of the minimal annihilating polynomial, $\Lambda_1$ is the assumed filter size used in constructing $\mathcal{T}(\hat f)$, and $\Lambda_2$ is the set of valid convolutions that satisfies $\Lambda_1\ast\Lambda_2 = \Gamma$. Note that $\mathcal{T}(\hat f)$ has dimensions $2|\Lambda_2|\times|\Lambda_1|$.}
\label{support_fig}
\end{figure}

We further assume that the edge-set of the image, specified by $E:=\cup_i \partial\Omega_i$, to be the zero-set of a 2-D bandlimited trigonometric polynomial:
\begin{equation}
\mu_0(\mbf r) = \sum_{\mbf k\in{ \Lambda_0}} c[\mbf k]\, e^{j2\pi\mbf k \cdot \mbf r},\quad \forall \mbf r\in{[0,1]}^2,
\label{eq:trigpoly}
\end{equation}
where the coefficients $c[\mbf k]\in\mathbb{C}$, and ${\Lambda_0}$ is a rectangular subset of $\Gamma$. Here we assume $\mu_0$ is the unqiue minimal degree trigonometric polynomial such that $E=\{\mu_0=0\}$, where the degree is defined by the rectangular dimensions of the Fourier support $\Lambda_0$. We have shown in \cite{siam} that when $\mu(\mbf r)$ is any bandlimited function that vanishes on the edgeset $E$, the gradient $\boldsymbol\nabla f = \left(\partial_x f, \partial_y f\right)$ satisfies the property 
\begin{equation}
\mu \boldsymbol\nabla f = 0,
\label{eq:spacedom2}
\end{equation}
in the distributional sense.  See Fig. \ref{illus} for an illustration when the number of regions $N=1$. Note that among all functions bandlimited to $\Lambda_0$ in Fourier domain, $\mu=\mu_0$ is the only one in this class that satisfies \eqref{eq:spacedom2}. However, if we consider $\mu$ that are bandlimited to a larger rectangular set $\Lambda_1$ with $\Lambda_0 \subset \Lambda_1 \subset \Gamma$, then we have shown that all $\mu$ satisfying \eqref{eq:spacedom2} are a multiple of $\mu_0$ \cite{siam}. The spatial domain relation \eqref{eq:spacedom2} translates directly to the following vector annihilation relation in the Fourier domain:
\begin{equation}
\sum_{\mbf k \in {\Lambda_1}} \widehat{\boldsymbol\nabla f}[\bs\ell - \mbf k]\;  \widehat{\mu}[\mbf k]= \bs 0, 
   ~~\forall ~\bs \ell\in \Lambda_2.
   \label{eq:annsys}
\end{equation}
Here $\widehat{\boldsymbol\nabla f}[\mbf k] = j2\pi(k_1 \widehat{f}[\mbf k],k_2 \widehat{f}[\mbf k])$ for $\mbf k=(k_1,k_2)$, and $\Lambda_2 \subset \mathbb{Z}^2$ is any rectangular set on which the convolutions between $\widehat{\boldsymbol\nabla f}$ and $\widehat{\mu}$ is valid. Note that the 2-D convolution between two filters support limited to rectangular sets $\Lambda_1$ and $\Lambda_2$ is support limited to the dilation of $\Lambda_1$ by $\Lambda_2$, which we denote by $\Lambda_1\ast\Lambda_2$. Since $\widehat{\mu}$ is bandlimited to $\Lambda_1$, when using samples of $\widehat{\boldsymbol\nabla f}$ within $\Gamma$, we require $\Lambda_1\ast\Lambda_2 = \Gamma$; see Fig.\ \ref{support_fig}. 

The Fourier domain annihilation relations \eqref{eq:annsys} can be compactly represented in matrix form as
\begin{equation}
\mathcal{T}({\hat f}) =
\begin{bmatrix}
\mathcal{T}_1({\hat f})\\
\mathcal{T}_2({\hat f})
\end{bmatrix} \mbf h = \bs 0 ,
\label{eq:annmatrix}
\end{equation}
where $\mathcal{T}_i(\widehat{f}) \in \mathbb C^{|\Lambda_2| \times |\Lambda_1|}$, $i=1,2$, are matrices corresponding to the discrete 2-D convolution of $k_i \widehat f[\mathbf k]$, $\mbf k \in \Gamma$, (omitting the irrelevant factor $j2\pi$) with a filter supported on $\Lambda_1$, with output restricted to the index set $\Lambda_2$. Here we use $\mbf h$ to denote the vectorized version of a filter $h[\mbf k], \mbf k \in \Lambda_1$. By our previous observations, the solutions $\mbf h$ to \eqref{eq:annmatrix} are given by the Fourier coefficients of a multiple of the minimal polynomial. Hence if the filter support $\Lambda_1$ is larger than the minimal filter support $\Lambda_0$, $\mathcal{T}({\hat f})$ has a large nullspace and is low-rank. Specifically, in \cite{siam} we proved the following:
\begin{prop}Suppose $\mathcal{T}(\widehat{f})$ is built with filter size $\Lambda_1\supseteq \Lambda_0$ satisfying $\Gamma \supseteq 2\Lambda_1\ast\Lambda_0$, then
\[
R:= \text{rank}~\mathcal{T}(\widehat{f}) = |\Lambda_1|-|\Lambda_1|\Lambda_0|
\] 
where $|\Lambda|$ is the number of indices in $\Lambda$ and $|\Lambda_1|\Lambda_0|$ is the number of integer shifts of $\Lambda_0$ contained in $\Lambda_1$. 
\end{prop}

\subsection{Recovery from non-uniform Fourier samples}
Since the matrix $\mathcal{T}(\widehat f)$ is low-rank, we propose to recover the signal from its noiseless Fourier samples as the convex optimization problem:
\begin{equation}
\label{nucnorm}
\mbox{minimize}_{\widehat g}~ \|\mathcal{T}(\widehat g)\|_* ~\mbox{subject to}~ \widehat{g}[\mbf k] =  \widehat{f}[\mbf k]; \mbf k\in \Theta
\end{equation}
where $\|\cdot\|_*$ denotes the nuclear norm, i.e. the sum of the singular values. To aid in our analysis, we will now reformulate the recovery of $\hat f$ as a matrix recovery problem using projection operators in the lifted matrix domain. We define basis matrices $\mbf A_{\mbf k} = \begin{bmatrix}\mbf A_{1,\mbf k}\\\mbf A_{2,\mbf k}\end{bmatrix} \in \mathbb{C}^{2|\Lambda_2|\times|\Lambda_1|}$, for all $\mbf k \in \Gamma$, where 
\begin{eqnarray}
(\mbf A_{i,\mbf k})_{\alpha,\beta} &=& \left\{\begin{array}{ccc}
\frac{k_i}{|\mbf k|\sqrt{|\omega_i(\mbf k)|}},&\mbox{ if }& (\alpha,\beta) \in \omega_i(\mbf k)\\
0 & \mbox{else}
\end{array}
\right.
\end{eqnarray}
for $i=1,2$. Here $\omega_i(\mbf k)$ is the set of locations of the matrix $\mathcal{T}(\widehat{f})$ containing copies of $k_i \widehat{f[\mbf k]}$.  Note that the set $\{\mbf A_{\mbf k}\}_{\mbf k \in \Gamma}$ forms an orthonormal basis for the space of structured matrices defined by the lifting $\mathcal{T}$. Specifically, for any set of coefficients $\{\widehat g[\mbf k]\}_{\mbf k \in \Gamma}$ we can expand the matrix $\mbf X = \mathcal{T}(\widehat g)$ as $\mbf X = \sum_{\mbf k} \inner{\mbf X,\mbf A_{\mbf k}} \mbf A_{\mbf k}$, where $\inner{\mbf X,\mbf A_{\mbf k}} = \widehat{g}[\mbf k]$. Using these basis matrices, we define the following operators in the lifted domain:
\begin{align}
\label{lifted}
\mathcal A_{\Theta}(\mbf X) &= \sum_{\mbf k\in \Theta} \inner{\mbf A_{\mbf k},\mbf X} \mbf A_{\mbf k}\\
\mathcal A^{\perp}(\mbf X) &= \mathcal I - \sum_{\mbf k \in \Gamma} \inner{\mbf A_{\mbf k},\mbf X} \mbf A_{\mbf k}\\
\label{Qomega}
\mathcal Q_{\Theta} &= \frac{|\Gamma|}{|\Theta|}\mathcal A_{\Theta} + \mathcal A^{\perp}
\end{align}
The constants in \eqref{Qomega} are chosen so that $\mbb E[\mathcal Q_{\Theta}] = \mathcal I$. 
Using these definitions, we rewrite \eqref{nucnorm} as
\begin{equation}
\label{stlr}
\mbox{minimize}_{\mbf X}~ \|\mbf X\|_* ~\mbox{subject to}~
 \mathcal Q_{\Theta}(\mbf X) = \mathcal  Q_{\Theta}(\mathcal{T}(\hat f))
\end{equation}

Several authors have shown that the performance of low-rank matrix recovery by nuclear norm minimization is dependent on the \emph{incoherence} of sampling basis with respect to the matrix to be to be recovered \cite{gross,chen2014robust}. Towards this end, we introduce two incoherence measures associated with $\mathcal{T}(\widehat f)$ defined solely in terms of the edge-set polynomial $\mu_0$. In the following, we set ${D}_{\Lambda_1}(\mbf r)$ to be the 2-D Dirichlet kernel supported on $\Lambda_1$, i.e. the function such that $\widehat{D_{\Lambda_1}}[\mbf k] = 1$ for all $\mbf k \in \Lambda_1$ and zero otherwise. For any collection of $N$ points $P = \{\mbf r_1,...,\mbf r_N\} \subset [0,1]^2$, we define the $N\times N$ Gram matrix $\mbf G(P)$ by
$(\mbf G(P))_{i,j} = D_{\Lambda_1}(\mbf r_i-\mbf r_j)$.
\begin{defn}\label{def:incoherence}
Let $\mu_0$ be a trigonometric polynomial bandlimited to $\Lambda_0$ (see \eqref{eq:trigpoly}), and set $R = |\Lambda_1|-|\Lambda_1|\Lambda_0|$. Define the incoherence measure $\rho_1$ by
\begin{equation}\label{eq:incoherence}
\frac{1}{\rho_1} = \max_{\substack{P \subset \{\mu_0=0\}\\|P|=R}} \sigma_{min}[\mbf G(P)]
\end{equation}
where $\sigma_{min}[\mbf G(P)]$ is the minimum singular value of $\mbf G(P)$.
\end{defn}
Put in words, among all possible arrangements of $R$ points along the edge-set $\{\mu_0=0\}$, we seek the arrangement that gives the best conditioning of the matrix $\mbf G(P)$, and call the resulting condition number $\rho_1$. Intuitively, the optimal arrangement will maximize the minimum separation distance among the $R$ points, and $\rho_1$ can be thought of as a measure of this geometric property. In particular, having any edges that enclose a small area will give a high $\rho_1$.

Additionally, our results rely on another incoherence measure related to properties of the gradient of $\mu_0$:.
\begin{defn}\label{def:incoherence}
Let $\mu_0$ be a trigonometric polynomial bandlimited to $\Lambda_0$ (see \eqref{eq:trigpoly}). Normalize $\mu_0$ such that $\int_{[0,1]^2} |\nabla\mu_0(\mbf r)|^2 d\mbf r = 1$. Define the incoherence measure $\rho_2$ by
\begin{equation}\label{eq:incoherence2}
\frac{1}{\rho_2} = \frac{\min_{\substack{\gamma\in B_{\Lambda_1}\\ \|\gamma\|_{2} = 1}} \int_{[0,1]^2} |\gamma(\mbf r)|^2|\nabla\mu_0(\mbf r)|^2 d\mbf r}{\|\widehat{\nabla\mu_0}\|^2_1}
\end{equation}
where $B_{\Lambda_1}$ denotes the space of all trigonometric polynomials bandlimited to $\Lambda_1$, and $\|\widehat{\nabla\mu_0}\|_1 = \sum_{\mbf k\in\Lambda_0}|\widehat{\nabla\mu_0}[\mbf k ]|$.
\end{defn}
Note that $\rho_2$ will be large when $|\nabla\mu_0|$ has several zeros, or equivalently, when $\mu_0$ has several critical points. Since $\mu_0$ must contain a critical point in every region defined by the complement of the edge-set, $\rho_2$ will be large when the image has several distinct regions. 


Now we state our main result:
\begin{thm}\label{mainthm}
Let $f$ be specified by \eqref{eq:pwc}, whose edge-set is described by the zero-set of the trigonometric polynomial $\mu_0$ bandlimited to $\Lambda_0$ (see \eqref{eq:trigpoly}) with associated incoherence measures $\rho_1$ and $\rho_2$. Let $\Theta\subset \Gamma$ be an index set drawn uniformly at random within $\Gamma$. Then there exists a universal constant $c>0$ such that the solution to \eqref{stlr} is exact ($\mbf X = \mbf \mathcal{T}(\widehat f)$) with probability exceeding $1-|\Gamma|^{-2}$, provided 
\begin{equation}\label{eq:numsamples}
|\Theta| > c\,\rho_1 \rho_2 R\,c_s\log^{4}|\Gamma|.
\end{equation}
where $R = \text{rank}~\mathcal{T}(\widehat{f})$ and $c_s = |\Gamma|/|\Lambda_1|$.
\end{thm}
Following the approach in \cite{chen2014robust}, we can prove this result by constructing an approximate dual certificate using the well-known golfing scheme of \cite{gross}. The adaptation of the proof in \cite{chen2014robust} to the measurement operator \eqref{Qomega} is straightforward, and these details are omitted for brevity. The essential difference of the above result and \cite{chen2014robust} is the characterization of the incoherency measures. The approach in \cite{chen2014robust} relies on an explicit low-rank factorization of the lifted matrix in terms of Vandermonde-like matricies, which is not available in our setting. Instead, we characterize the row and column spaces of the matrix and use it to prove the above result.
\subsection{Row and column spaces of $\mathcal{T}(\hat f)$ and incoherence}\label{sec:incoherence}
Define $\mathcal P_U$ and $\mathcal P_V$ to be the orthogonal projections onto the column space and row space of $\mathcal{T}(\hat f)$, respectively, i.e.\ if $\mathcal{T}(\widehat{f}) = \mbf U\mbf \Sigma\mbf V^*$ is the rank-$R$ singular value decomposition then $\mathcal P_U \mbf X = \mbf U\mbf U^* \mbf X$, $\mathcal P_V \mbf X = \mbf X \mbf V\mbf V^*$. One can show it is possible to construct an approximate dual certificate with high probability \cite{chen2014robust,gross}, provided we can uniformly bound the norms of the projections $\mathcal{P}_U\mbf A_{\mbf k}$ and $\mathcal{P}_V\mbf A_{\mbf k}$. The following proposition shows these norms can be controlled by the incoherence measures introduced in \eqref{eq:incoherence} and \eqref{eq:incoherence2}

\begin{prop}\label{prop:coherence}
Consider $\mathcal{T}(\hat f)$ of rank $R$ corresponding to a piecewise constant function $f$ whose edge set coincides with the zero set of $\mu_0$, let $\rho_1$ and $\rho_2$ be the incoherency measures of $\mu_0$, and set $c_s = |\Gamma|/|\Lambda_1|$. Then we have
\begin{equation}
\label{coherence2}
\max_{\mbf k \in \Gamma}\{\|\mathcal P_{U}\mathbf A_{\mbf k}\|,\|\mathcal P_{V}\mathbf A_{\mbf k}\} \|_F^2 \leq \frac{\rho_1 \rho_2 R c_s}{|\Gamma|} 
\end{equation}
\end{prop}
The proof relies on the following basis representations for the row and column spaces of $\mathcal{T}(\widehat{f})$:
\begin{lem}\label{lem:rightnull}
Choose any $N \geq |\Lambda|+|\Lambda_0|-|\Lambda|\Lambda_0|$ points $\{\mbf r_1,...,\mbf r_N\} \subset \{\mu_0 = 0\}$, and define $B$ to be the collection of filters $\{\mbf h_1,...,\mbf h_N\}$ where each $\mbf h_i \in \mathbb{C}^{|\Lambda|}$ are the Fourier coefficients of the translated Dirichlet kernel $D_{\Lambda_1}(\mbf r-\mbf r_i)$. Then there exists a subset $R = \text{rank}~\mathcal{T}(\widehat{f})$ elements from $B$ that is a basis for the row space of $\mathcal{T}(\widehat{f})$.
\end{lem}

\begin{lem}\label{lem:leftnull}
Let $B_R = \{\mbf d_i\}_{i=1}^R$ be any basis of the row space of $\mathcal{T}(\widehat f)$, and set $\widehat{\nabla \mu_0} = (\mbf c_x,\mbf c_y)\in \mathbb{C}^{2|\Lambda_0|}$. Then $B_L = \{(\mbf d_i\ast \mbf c_x, \mbf d_i \ast \mbf c_y)\}_{i=1}^R$ is a basis of the column space of $\mathcal{T}(\widehat f)$.
\end{lem}

\begin{figure}[ht!]
\centering
\begin{minipage}{0.315\linewidth}
\vspace{-1em}
\begin{tikzpicture}
    \node[anchor=south west,inner sep=0] (image) at (0,0) {\includegraphics[width=\linewidth, trim = 150 50 100 150,clip]{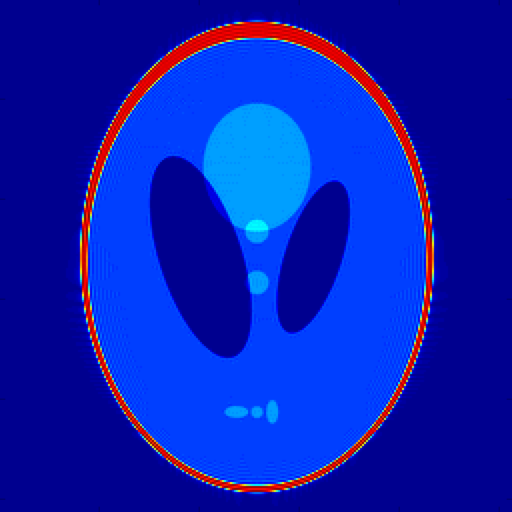}}; 
    \node[anchor=north east,inner sep=0, draw, white, thick] (image2) at (2.75,1.25) {\includegraphics[height=12.5mm,width=12.5mm]{images/lowrank_SL_fullsamp_ifft_jet.png}};
    \begin{scope}[x={(image.south east)},y={(image.north west)}]
        \draw[yellow,thick] (0.68,0.02) rectangle (0.90,0.22);
    \end{scope}    
\end{tikzpicture}

{\centering \footnotesize Fully sampled\\ }
\end{minipage}
\begin{minipage}{0.315\linewidth}
\begin{tikzpicture}
    \node[anchor=south west,inner sep=0] (image) at (0,0) {\includegraphics[width=\linewidth, trim = 150 50 100 150,clip]{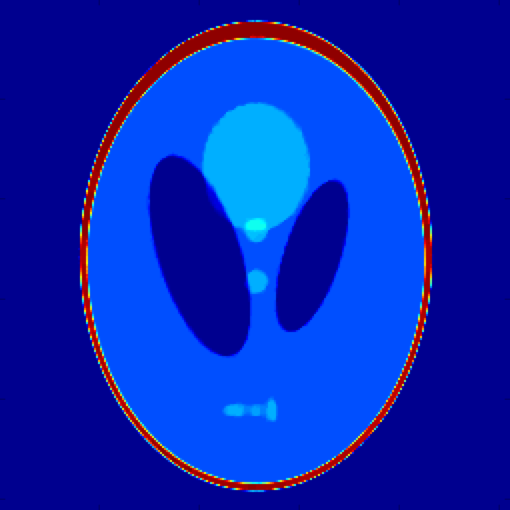}}; 
    \node[anchor=north east,inner sep=0, draw, white, thick] (image2) at (2.75,1.25) {\includegraphics[height=12.5mm,width=12.5mm]{images/lowrank_SL_vardensity0p05_nonoise_TV_ifft_jet.png}};
    \begin{scope}[x={(image.south east)},y={(image.north west)}]
        \draw[yellow,thick] (0.68,0.02) rectangle (0.90,0.22);
    \end{scope}    
\end{tikzpicture}

{\centering \footnotesize TV regularized recovery\\ }
\end{minipage}
\begin{minipage}{0.315\linewidth}
\begin{tikzpicture}
    \node[anchor=south west,inner sep=0] (image) at (0,0) {\includegraphics[width=\linewidth, trim = 150 50 100 150,clip]{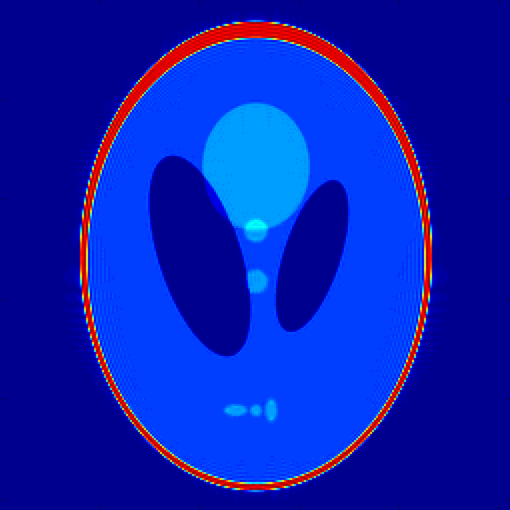}}; 
    \node[anchor=north east,inner sep=0, draw, white, thick] (image2) at (2.75,1.25) {\includegraphics[height=12.5mm,width=12.5mm]{images/lowrank_SL_vardensity0p05_nonoise_nucUV_1000iter_ifft_jet.png}};
    \begin{scope}[x={(image.south east)},y={(image.north west)}]
        \draw[yellow,thick] (0.68,0.02) rectangle (0.90,0.22);
    \end{scope}    
\end{tikzpicture}

{\centering \footnotesize Structured low-rank recovery\\ }
\end{minipage}
\caption{\small Recovery of synthetic MRI data from 20-fold variable density undersampling.}
\label{fig:SLresult}
\end{figure}


\begin{figure}[ht!]
\centering
\begin{minipage}{0.315\linewidth}
\vspace{-0.2em}
\begin{tikzpicture}
    \node[anchor=south west,inner sep=0] (image) at (0,0) {\includegraphics[height=\linewidth,width=0.9\linewidth,trim=60 70 150 70,clip,angle=-90,origin=c]{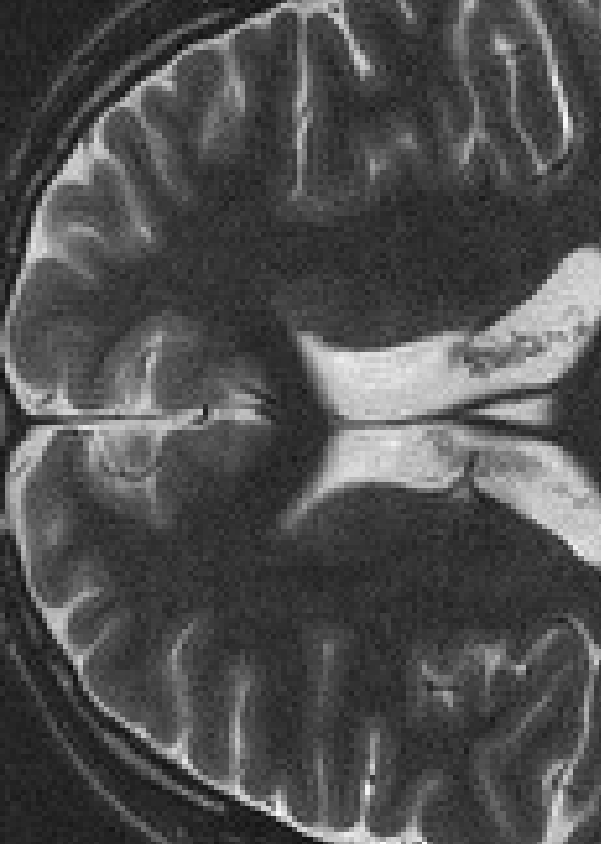}};
    \begin{scope}[x={(image.south east)},y={(image.north west)}]
        \draw[yellow,thick,-latex] (0.25,0.75) -- (0.40,0.80);
        \draw[yellow,thick,-latex] (0.42,0.2) -- (0.27,0.2);
    \end{scope}    
    \node[anchor=north east,inner sep=0, draw, white, thick] (image2) at (2.75,2.60) {\includegraphics[height=10mm,width=12.5mm,angle=-90,origin=c]{images/realbrain_cor2_hi_uniform0p5_redo_orig_gray.pdf}};
    \begin{scope}[x={(image.south east)},y={(image.north west)}]
        \draw[cyan,thick] (0.70,0.71) rectangle (0.95,0.94);
    \end{scope}    
\end{tikzpicture}
\centering
\includegraphics[width=0.9\linewidth,height=0.9\linewidth,angle=-90,origin=c]{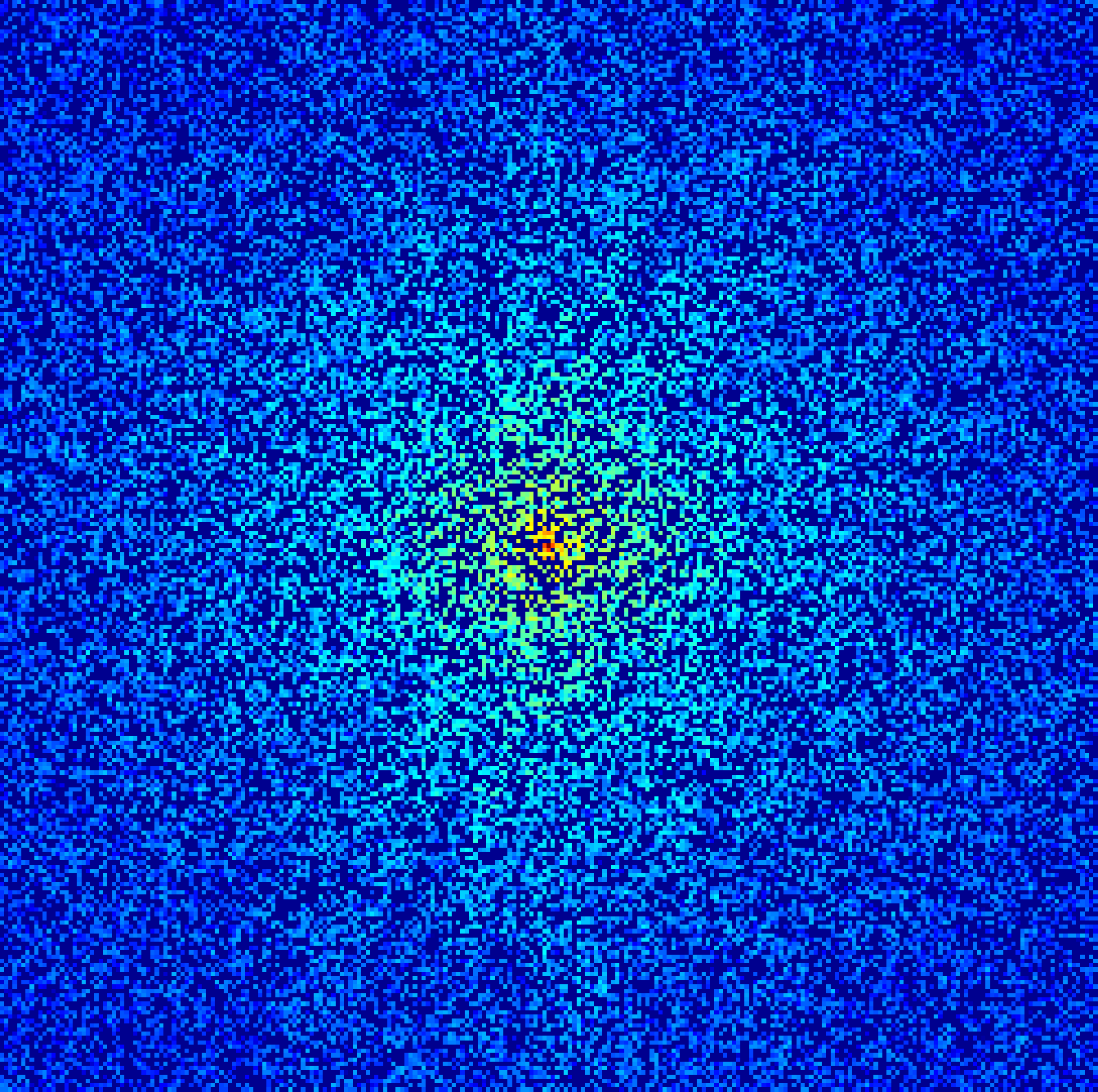}

{\centering \footnotesize Fully sampled (top)\\Subsampled $k$-space (bottom)\\ }
\end{minipage}
\begin{minipage}{0.315\linewidth}
\begin{tikzpicture}
    \node[anchor=south west,inner sep=0] (image) at (0,0) {\includegraphics[height=\linewidth,width=0.9\linewidth,trim=60 70 150 70,clip,angle=-90,origin=c]{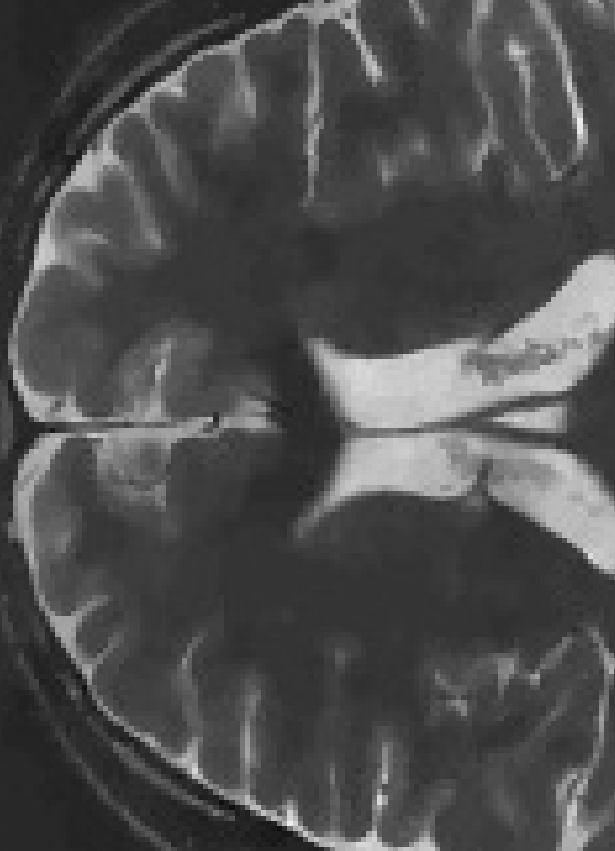}};
    \begin{scope}[x={(image.south east)},y={(image.north west)}]
        \draw[yellow,thick,-latex] (0.25,0.75) -- (0.40,0.80);
        \draw[yellow,thick,-latex] (0.42,0.2) -- (0.27,0.2);
    \end{scope}    
    \node[anchor=north east,inner sep=0, draw, white, thick] (image2) at (2.75,2.60) {\includegraphics[height=10mm,width=12.5mm,angle=-90,origin=c]{images/realbrain_cor2_hi_uniform0p5_redo_tv_gray.pdf}};    
    \begin{scope}[x={(image.south east)},y={(image.north west)}]
        \draw[cyan,thick] (0.70,0.71) rectangle (0.95,0.94);
    \end{scope}        
\end{tikzpicture}
\begin{tikzpicture}
    \node[anchor=south west,inner sep=0] (image) at (0,0) {\includegraphics[height=\linewidth,width=0.9\linewidth,trim=60 70 150 70,clip,angle=-90,origin=c]{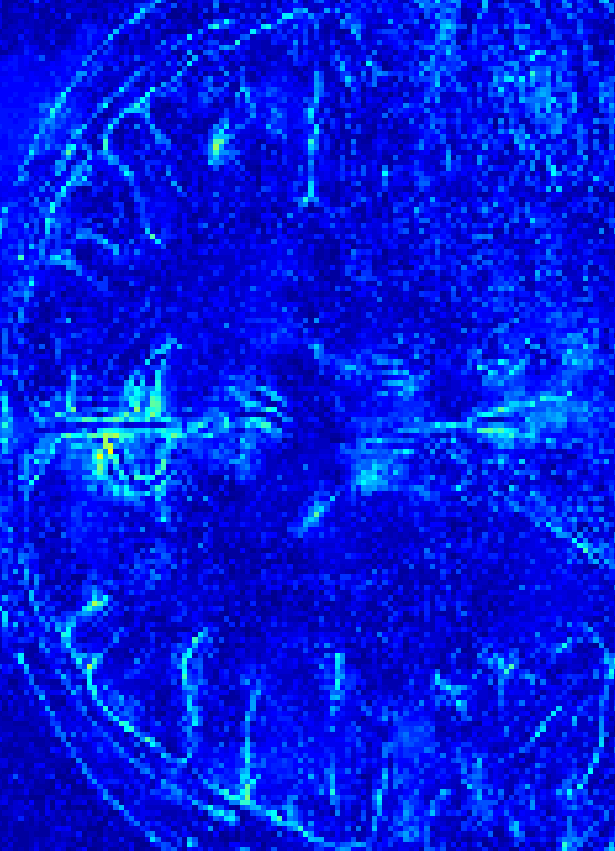}};
    \begin{scope}[x={(image.south east)},y={(image.north west)}]
        \draw[yellow,thick,-latex] (0.25,0.75) -- (0.40,0.80);
        \draw[yellow,thick,-latex] (0.42,0.2) -- (0.27,0.2);
    \end{scope}
\end{tikzpicture}

{\centering \footnotesize TV regularized recovery\\SNR=17.8dB\\ }
\end{minipage}
\begin{minipage}{0.315\linewidth}
\begin{tikzpicture}
    \node[anchor=south west,inner sep=0] (image) at (0,0) {\includegraphics[height=\linewidth,width=0.9\linewidth,trim=60 70 150 70,clip,angle=-90,origin=c]{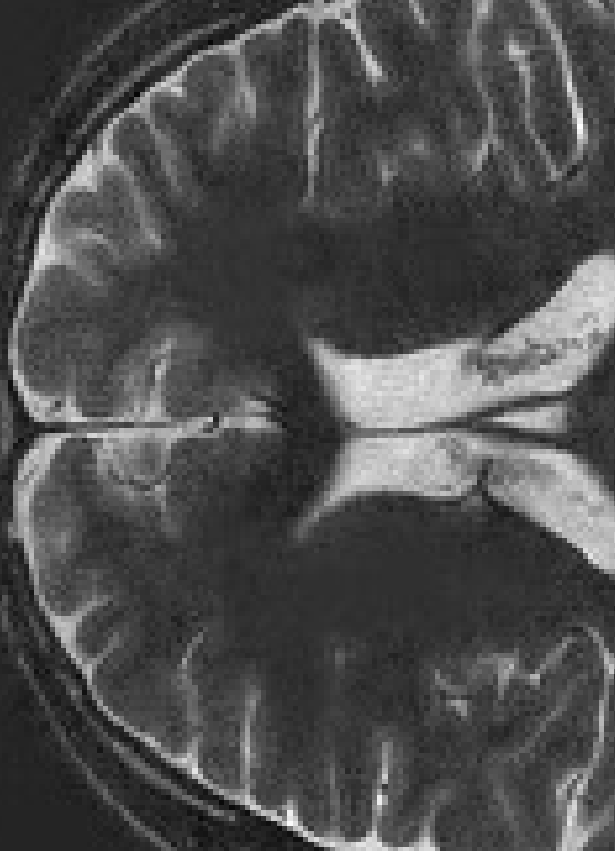}};
    \begin{scope}[x={(image.south east)},y={(image.north west)}]
        \draw[yellow,thick,-latex] (0.25,0.75) -- (0.40,0.80);
        \draw[yellow,thick,-latex] (0.42,0.2) -- (0.27,0.2);
    \end{scope}
    \node[anchor=north east,inner sep=0, draw, white, thick] (image2) at (2.75,2.60) {\includegraphics[height=10mm,width=12.5mm,angle=-90,origin=c]{images/realbrain_cor2_hi_uniform0p5_redo_lrg_gray.pdf}};    
    \begin{scope}[x={(image.south east)},y={(image.north west)}]
        \draw[cyan,thick] (0.70,0.71) rectangle (0.95,0.94);
    \end{scope}        
\end{tikzpicture}
\begin{tikzpicture}
    \node[anchor=south west,inner sep=0] (image) at (0,0) {\includegraphics[height=\linewidth,width=0.9\linewidth,trim=60 70 150 70,clip,angle=-90,origin=c]{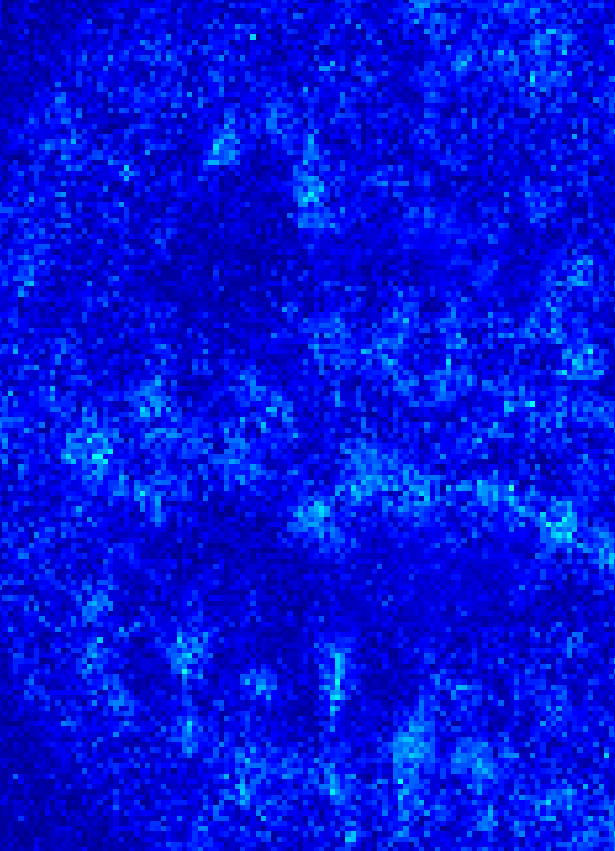}};
    \begin{scope}[x={(image.south east)},y={(image.north west)}]
        \draw[yellow,thick,-latex] (0.25,0.75) -- (0.40,0.80);
        \draw[yellow,thick,-latex] (0.42,0.2) -- (0.27,0.2);
    \end{scope}
\end{tikzpicture}

{\centering \footnotesize Structured low-rank recovery\\SNR=19.0dB\\ }
\end{minipage}
\caption{\small Recovery of real MRI data from 2-fold random uniform undersampling. Error images shown below.}
\label{fig:mainresult}
\end{figure}

\section{EXPERIMENTS}

In Fig.\ \ref{fig:SLresult} we demonstrate the recovery of a synthetic piecewise constant phantom \cite{guerquin2012realistic} ($181\times181$ sampling grid, single channel) from 20-fold variable density random Fourier samples using the structured low-rank matrix completion approach \eqref{nucnorm}. We solve \eqref{nucnorm} using a singular value thresholding approach proposed in \cite{sampta2015}. The filter size was set to $33\times25$. Compared with total variation (TV) minimization, the proposed structured low-rank approach more accurately recovers the original piecewise constant regions.

Additionally, in Fig.\ \ref{fig:mainresult} we demonstrate the structured low-rank approach on the recovery real MR data ($255$$\times$$255$ sampling grid with 4 receiver coils, coil-compressed to a single channel) from 2-fold uniform random undersampling, using a filter size of $45\times45$. Due to the problem size, the formulation \eqref{nucnorm} is difficult to solve via singular value thresholding. Instead, we make use of the recently proposed GIRAF algorithm \cite{isbi2016} which solves an approximated version of \eqref{nucnorm}. The result shows similar benefit over a TV regularized recovery in its ability to preserve fine details and strong edges.

\section{DISCUSSION AND CONCLUSION}
We derived a performance guarantee for the recovery of piecewise constant images from non-uniform Fourier samples by a structured matrix completion. This was achieved by adapting results in \cite{chen2014robust} to the case of a low-rank multifold Toeplitz structure with an additional weighting scheme. We also define new incoherence measures that rely only on properties of the minimal annihilating polynomial whose zero-set encodes the edges of the image.

While in the present work we only consider noiseless ideal samples, in future work we intend to derive guarantees for robust recovery in the presence of noise and model-mismatch.  Additionally, it would be interesting to adapt our results to a wider variety of sampling distributions, and to identify the optimal sampling strategy for signals belonging to our image model.
\newpage
\bibliographystyle{IEEEtran}
\bibliography{root}

\end{document}

%% file: definitions.tex
\newcommand{\mbf}[1]{{\mathbf #1}}
\newcommand{\mbb}{\mathbb}
\newcommand{\bs}{\boldsymbol}

\newcommand{\inner}[1]{\left\langle #1\right\rangle}
